\documentclass[conference]{IEEEtran}

\IEEEoverridecommandlockouts

\usepackage{cite}
\usepackage{amsmath,amssymb,amsfonts}
\usepackage{algpseudocode}
\usepackage{graphicx}
\usepackage{pseudo}
\usepackage{textcomp}
\usepackage{xcolor}
\def\BibTeX{{\rm B\kern-.05em{\sc i\kern-.025em b}\kern-.08em
    T\kern-.1667em\lower.7ex\hbox{E}\kern-.125emX}}

\usepackage{balance}
\usepackage{multirow}
\usepackage{color}

\begin{document}

\title{Parallel Vertex Cover Algorithms on GPUs
\thanks{This work is supported by the Mamdouha El-Sayed Bobst Deanship Fund in the Faculty of Arts and Sciences at the American University of Beirut.}
}

\author{
\IEEEauthorblockN{Peter Yamout, Karim Barada, Adnan Jaljuli, Amer E. Mouawad, Izzat El Hajj}\\
\IEEEauthorblockA{\textit{American University of Beirut, Lebanon}}
}

\maketitle

\begin{abstract}

Finding small vertex covers in a graph has applications in numerous domains such as scheduling, computational biology, telecommunication networks, artificial intelligence, social science, and many more.
Two common formulations of the problem include: Minimum Vertex Cover (MVC), which finds the smallest vertex cover in a graph, and Parameterized Vertex Cover (PVC), which finds a vertex cover whose size is less than or equal to some parameter $k$.
Algorithms for both formulations involve traversing a search tree, which grows exponentially with the size of the graph or the value of $k$.

Parallelizing the traversal of the vertex cover search tree on GPUs is challenging for multiple reasons.
First, the search tree is a narrow binary tree which makes it difficult to extract enough sub-trees to process in parallel to fully utilize the GPU's massively parallel execution resources.
Second, the search tree is highly imbalanced which makes load balancing across a massive number of parallel GPU workers especially challenging.
Third, keeping around all the intermediate state needed to traverse many sub-trees in parallel puts high pressure on the GPU's memory resources and may act as a limiting factor to parallelism.

To address these challenges, we propose an approach to traverse the vertex cover search tree in parallel using GPUs while handling dynamic load balancing.
Each thread block traverses a different sub-tree using a local stack, however, we use a global worklist to balance the load to ensure that all blocks remain busy.
Blocks contribute branches of their sub-trees to the global worklist on an as-needed basis, while blocks that finish their sub-trees pick up new ones from the global worklist.
We use degree arrays to represent intermediate graphs so that the representation is compact in memory to avoid limiting parallelism, but self-contained which is necessary for the load balancing process.

Our evaluation shows that compared to approaches used in prior work, our hybrid approach of using local stacks and a global worklist substantially improves performance and reduces load imbalance, especially on difficult instances of the problem.
Our implementations have been open sourced to enable further research on parallel solutions to the vertex cover problem and other similar problems involving parallel traversal of narrow and highly imbalanced search trees.

\end{abstract}

\section{Introduction}\label{sec-intro}

A vertex cover of a graph is a set of vertices whose deletion from the graph (along with incident edges) induces an edgeless graph.
Finding small vertex covers is one of the most famous problems in algorithmic graph theory and is among the original 21 NP-complete problems introduced by Karp in 1972~\cite{DBLP:conf/coco/Karp72}.
Finding small vertex covers has many applications in numerous domains such as scheduling, computational biology, telecommunication networks, artificial intelligence, social science, and many more~\cite{895327,article-vc-app}.
It is a problem that is particularly well-studied from the view point of parameterized complexity~\cite{DBLP:journals/tcs/ChenKX10,DBLP:journals/ipl/BalasubramanianFR98}, kernelization~\cite{DBLP:conf/birthday/FellowsJKRW18,DBLP:reference/algo/Chen08a}, approximation~\cite{DBLP:journals/talg/Karakostas09,DBLP:conf/iwoca/DelbotLP13}, exact exponential-time algorithms~\cite{DBLP:reference/algo/Grandoni16,DBLP:conf/aussois/Woeginger01}, and heuristics~\cite{DBLP:conf/eps/Evans98,DBLP:journals/heuristics/VossF12}.

We consider two common formulations of the problem: {\sc Minimium Vertex Cover} (MVC), which finds a vertex cover with the smallest number of vertices, and {\sc Parameterized Vertex Cover} (PVC) which finds a vertex cover with $\leq k$ vertices for a given integer $k > 0$.
Most algorithms for MVC and PVC traverse a binary tree to search for vertex covers, following the branch-and-reduce paradigm, also commonly known as branch-and-bound.
At each node in the tree, reduction rules are first applied, followed by a check if a stopping criteria has been reached or if a solution has been found.
If neither is the case, the tree branches into two sub-problems: one that removes the highest-degree vertex from the graph and adds it to the solution, and another that removes the neighbors of the highest-degree vertex from the graph and adds them to the solution.

Parallelizing the traversal of the vertex cover search tree on GPUs comes with many challenges.
One challenge is extracting enough parallelism to fully utilize the GPU.
Prior works~\cite{kabbara2013parallel,abu2018accelerating} divide the tree into sub-trees starting at the same depth and distribute these sub-trees across thread blocks.
Since the search tree is narrow (binary), the sub-trees need to start at a deep level to ensure that enough parallelism is extracted.
However, the deeper the starting level, the higher the overhead incurred for reaching these sub-trees due to redundancy~\cite{abu2018accelerating} or grid launches and memory consumption~\cite{kabbara2013parallel}.
Another challenge is dealing with load imbalance.
Load imbalance is particularly challenging because the vertex cover search tree is highly imbalanced, so sub-trees have dramatically different sizes.
Load imbalance on GPUs is typically addressed by extracting even more parallelism than the number of thread blocks that can run simultaneously to allow for dynamic load balancing.
Indeed, prior work~\cite{abu2018accelerating} extracts many more sub-trees than thread blocks for this reason.
However, the imbalance in the vertex cover search tree is so high that one would need to start at very deep levels to ensure adequate load balancing, thus further increasing the overhead of reaching these sub-trees.
A third challenge is ensuring that memory does not become a limiting factor for parallelism.
As each thread block traverses a sub-tree, it needs to reserve a large amount of memory to maintain the intermediate traversal state.
Hence, the memory capacity can limit the number of thread blocks that can execute in parallel.

One way to address these challenges is to use a global worklist that dynamically distributes work across thread blocks on a per-tree-node instead of a per-sub-tree basis.
However, such an approach would result in high contention on the queue and an exponential explosion in the number of queue entries.
Instead, we propose a hybrid approach where each thread block uses a local stack to traverse a sub-tree, but contributes branches of its sub-tree to a global worklist as needed to ensure that there is enough work to keep all thread blocks busy.
This hybrid approach extracts just enough parallelism to ensure load balance without incurring the overhead of redundancy or grid launches and memory capacity.
We represent the intermediate graphs using degree arrays to ensure that the representation is compact so that memory consumption does not limit parallelism, but at the same time self-contained so that intermediate graphs can be shared across different thread blocks in the load balancing process.

We implement CUDA kernels for solving both MVC and PVC using our proposed approach and evaluate them on a server-grade GPU.
Our evaluation shows that compared to the approach of distributing sub-trees starting at the same depth across thread blocks, our hybrid approach substantially improves performance and reduces load imbalance, especially on difficult instances of the problem and on graphs with a high average degree.

\section{Background}\label{sec-background}

\subsection{Vertex Cover}

We assume a graph $G = (V, E)$ that is finite, simple, and undirected~\cite{DBLP:books/daglib/0030488}.
The \emph{neighborhood} of a vertex $v \in V(G)$ is the set of vertices adjacent to $v$, denoted by $N_G(v)$.
The \emph{degree} of a vertex $v \in V(G)$ is the number of edges incident on $v$, denoted by $d_G(v)$.
The subscript $G$ will be dropped when clear from the context.
The maximum degree in G is denoted by $\Delta(G)$.
Given a set of vertices $S \subseteq V(G)$, the subgraph induced by $S$ is denoted by $G[S]$.
$G - v$ and $G - S$ denote $G[V(G) \setminus \{v\}]$ and $G[V(G) \setminus S]$, respectively.

A set $S \subseteq V(G)$ is a \emph{vertex cover} for $G$, if for every edge $uv \in E(G)$, we have $\{u, v\} \cap S \neq \emptyset$.
Alternatively, one can view a vertex cover as a set of vertices whose deletion from the graph (along with incident edges) induces an edgeless graph.
A vertex cover $S$ is a \emph{minimum vertex cover} for $G$ if there is no vertex cover $S'$ for $G$ such that $|S'| < |S|$.

We consider two common formulations of the problem of finding small vertex covers in graphs.
One formulation is {\sc Minimium Vertex Cover} (MVC) which aims to find the minimum vertex cover $S$ of $G$.
Another formulation is {\sc Parameterized Vertex Cover} (PVC) which, for a given integer $k > 0$, aims to find a vertex cover $S$ of $G$ such that $|S| \leq k$, if such a vertex cover exists.
When $k$ is larger than (or equal to) the size of a minimum vertex cover, PVC tends to be ``faster'' than MVC because the search terminates as soon as a solution of size $k$ or less is found, in contrast with MVC where the search continues exploring the solution space to guarantee that no smaller solution exists. 

\subsection{Algorithms for Finding Vertex Covers}\label{sec:background-alg}

Most algorithms for MVC and PVC follow the well-known branch-and-reduce paradigm~\cite{DBLP:conf/aussois/Woeginger01}.
A \emph{branch-and-reduce} algorithm searches the complete solution space of a given problem by \emph{branching}, i.e., making decisions and solving smaller sub-problems.
Due to the exponentially increasing number of potential solutions, the solution space is pruned using \emph{reduction rules} derived from bounds on the function to be optimized and/or the value of the current best solution.
At the implementation level, branch-and-reduce algorithms translate to search-tree-based algorithms.
The search tree size usually grows exponentially with either the size of the input or, in the parameterized version, the value of the parameter $k$.

\begin{figure}[t]
    \small
    \include{fig/2-background/serial-mvc}
    \vspace{-10pt}
    \caption{The serial algorithm for {\sc Minimum Vertex Cover}. Initially $S = \emptyset$ and we assume that the graph has at least one edge, i.e., $|E(G)| \geq 1$.}\label{fig:serial-mvc}
\end{figure}

\begin{figure}
    \centering
    \includegraphics[width=0.9\columnwidth]{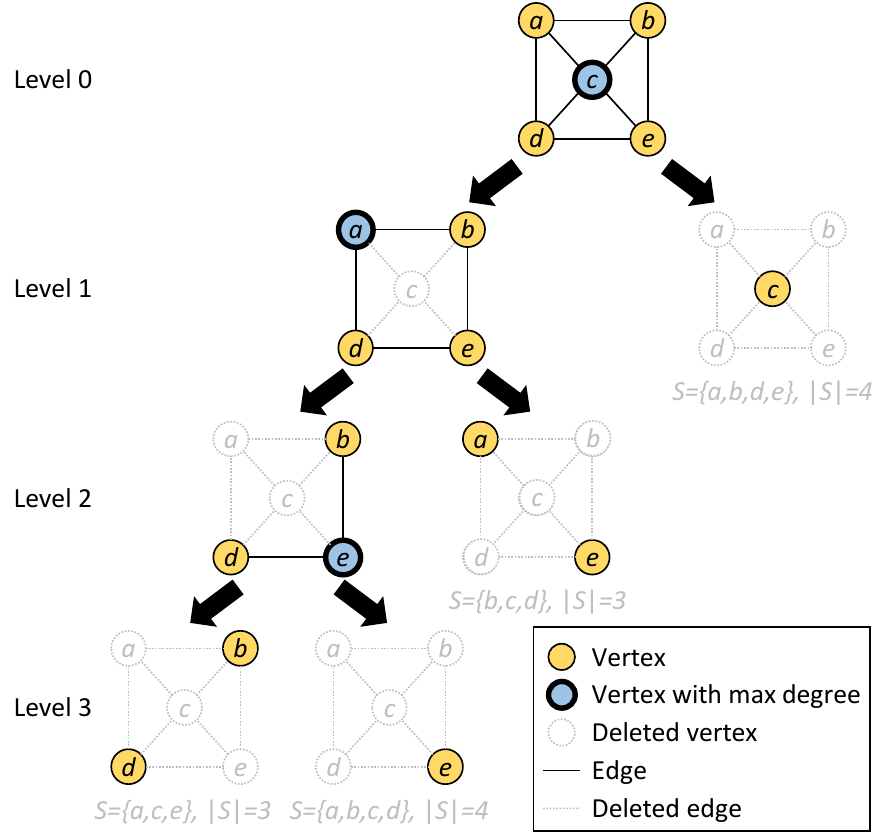}
    \caption{Example of Vertex Cover Search Tree}\label{fig:mvc-tree-example}
\end{figure}

Figure~\ref{fig:serial-mvc} shows a branch-and-reduce algorithm for solving the MVC problem by traversing the vertex cover search tree.
Figure~\ref{fig:mvc-tree-example} shows the tree traversed for an example graph (reduction rules are not applied in the example to keep the example small).
The search tree is binary, branching into two sub-problems at every node in the tree.
When visiting a node, the reduction rules are first applied (line 4, described later).
Next, a stopping condition is checked to see if the search should stop or if it is still possible to find a better solution at this node or its descendants (line 5, described later).
If it is possible, the algorithm checks if it has already arrived at a vertex cover (line 7) and updates the best solution accordingly (line 8).
If it has not arrived at a vertex cover, then it needs to branch.
A vertex in the graph is selected as the basis for branching, which is typically the vertex of highest degree denoted by  $v_{max}$ (line 10).
One branch removes $v_{max}$ from the graph and adds it to the solution, while the other branch removes all the neighbors of $v_{max}$ from the graph and adds them to the solution.
The MVC function is called recursively to try and find a solution on each branch.

To initialize $\textsf{best}$ (line 1), the minimum vertex cover is approximated using a greedy algorithm.
The algorithm applies all reduction rules to the graph, removes the largest degree vertex from the graph (hence adding it to a solution), and repeats this process until a vertex cover is found.

The reduction rules applied are the \emph{degree-one} reduction rule, the \emph{degree-two-triangle} reduction rule, and the \emph{high-degree} reduction rule.
The rules are repeatedly applied until the graph stops changing.
The degree-one reduction rule (lines 17-19) states that for any vertex $v$ of degree one, either $v$ or its neighbor $u$ need to be in a solution to cover the edge $uv$.
It is always at least as good to include $u$ as to include $v$ because $u$ may have other incident edges that would also be covered.
For the degree-two-triangle reduction rule (lines 21-23), if $G$ contains a vertex $v$ such that $N(v) = \{u,w\}$ and $uw \in E(G)$ (i.e., $u$, $v$, and $w$ form a triangle), then two of the three vertices are needed to cover all the edges in the triangle.
It is always at least as good to include $u$ and $w$ as to include only one of them and $v$ because $u$ and $w$ may have other incident edges that would also be covered.
Finally, the high-degree rule states that whenever a vertex $v$ is found whose degree is greater than $\textsf{best} - |S| - 1$, then adding all the neighbors of $v$ to $S$ can never achieve a solution better than $\textsf{best}$.
Therefore, $v$ is added to the solution.

The stopping condition (line 5) identifies if it is possible to find a solution at a node in the tree or its descendants.
The condition deems it impossible in one of two sub-conditions.
The first sub-condition is if the number of vertices added to a solution so far already exceeds $\textsf{best}$.
The second sub-condition is based on the observation that the high-degree reduction rule has already removed all vertices with degree $> \textsf{best} - |S| - 1$.
Hence, the remaining vertices have degree at most $\textsf{best} - |S| - 1$.
Moreover, finding a solution better than $\textsf{best}$ would entail including no more than $\textsf{best} - |S| - 1$ vertices.
Hence, the maximum number of edges that can be covered to find such a solution is $(\textsf{best} - |S| - 1)^2$.
If the graph has more than that number of edges, then it is impossible to find a better solution on that branch.

As for the PVC problem, we omit the pseudocode for space constraints because it is largely similar to the pseudocode for MVC in Figure~\ref{fig:serial-mvc} with a few differences.
For the high-degree reduction rule, $k - |S|$ is used instead of $\textsf{best} - |S| - 1$ for comparison.
For the stopping condition, the number of deleted vertices is compared to the parameter $k$ instead of $\textsf{best}$, and the number of edges is compared to $(k - |S|)^2$ instead of $(\textsf{best} - |S| - 1)^2$.
When a vertex cover is found that does not exceed $k$, rather than updating $\textsf{best}$ and continuing, the search is ended.

\section{Challenges}\label{sec:challenges}

Parallelizing the traversal of the vertex cover search tree on GPUs comes with numerous challenges.
We discuss some of these challenges in this section and discuss how prior work has addressed them.

\subsection{Challenge \#1: Extracting Massive Parallelism}

GPUs are massively parallel processors which require thousands of threads to fully utilize their computational resources.
Hence, one must be able to extract many units of independent work from an application to parallelize the application on the GPU effectively.
The typical way of extracting parallelism from a search tree traversal is by traversing independent sub-trees in parallel.
For some graph applications that require search tree traversal~\cite{jenkins2011lessons,almasri2021k}, the search tree is wide because the branching factor is large, which means that enough independent sub-trees can be extracted at the first or second level of the tree.
However, the vertex cover search tree is a narrow binary tree, which means that enough parallelism is not available until deeper levels of the tree.

\begin{figure}
    \centering
    \includegraphics[width=1.0\columnwidth]{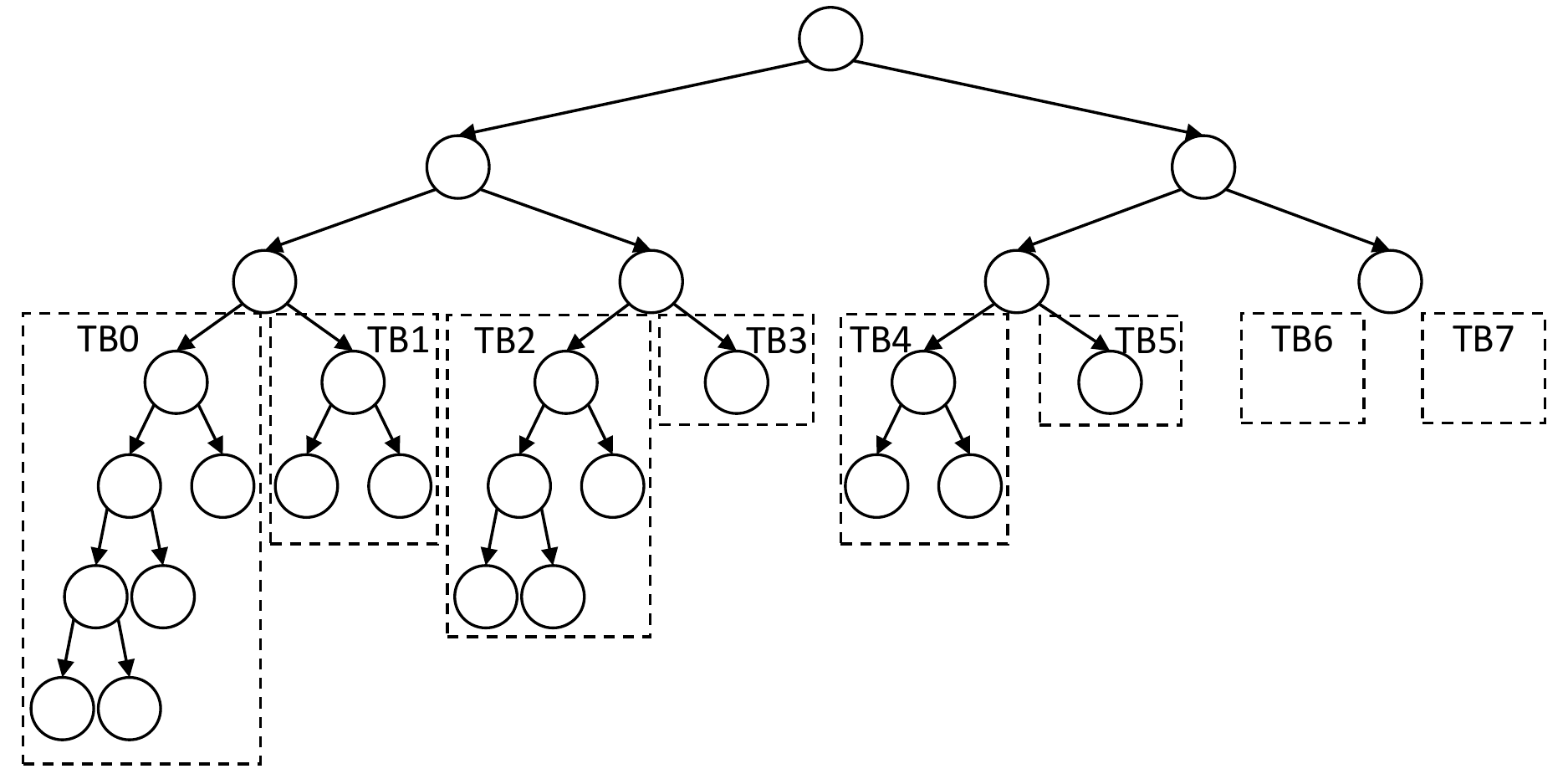}
    \caption{Traversing Sub-trees in Parallel by Different Thread Blocks}\label{fig:parallel-sub-trees}
\end{figure}

Prior work on accelerating vertex cover on GPUs~\cite{abu2018accelerating,kabbara2013parallel} consider a specific depth of the tree as the starting level, and treat all sub-trees starting at that level as independent units of parallelism.
Sub-trees are distributed across thread blocks and each thread block traverses its sub-tree in a depth-first manner.
This approach is illustrated in Figure~\ref{fig:parallel-sub-trees}. 
To reach the sub-trees, one approach~\cite{kabbara2013parallel} is to consecutively launch a separate grid for each level until the starting level is reached.
However, this approach requires launching multiple grids and storing the state of all the initial sub-trees simultaneously.
The deeper the starting level, the more the grid launches needed and the more the memory needed to store the state of the initial sub-trees.
Hence, there is a trade-off between the amount of parallelism extracted, and the grid launch and memory capacity overhead incurred by the extraction process.
Another approach~\cite{abu2018accelerating} is for each thread block to make its way down to its sub-tree from the root.
However, this approach causes thread blocks assigned to nearby sub-trees to take redundant steps to arrive to their sub-trees.
For example, TB0 and TB1 in Figure~\ref{fig:parallel-sub-trees} both visit the same tree nodes in the first three levels.
The deeper the starting level, the more the redundant work performed by different thread blocks.
Hence, there is a trade-off between the amount of parallelism extracted and the amount of redundant work performed.

\subsection{Challenge \#2: Load Balancing}\label{sec:challenge-balance}

The massively parallel nature of GPUs makes them particularly sensitive to load imbalance.
This sensitivity is especially challenging for the vertex cover problem.
The vertex cover search tree is highly imbalanced because whenever the tree branches, one branch removes a single high-degree vertex from the graph whereas the other branch removes all the neighbors of that high-degree vertex; hence, the latter branch is likely to exceed the current minimum and terminate sooner.
Because the search tree is imbalanced, prior work's approach of extracting parallelism via sub-trees starting at the same level leads to high load imbalance.
For example, in Figure~\ref{fig:parallel-sub-trees}, TB0 receives a large sub-tree, TB5 only receives a single node, and TB7 does not even have a sub-tree to traverse.

One way to mitigate load imbalance on GPUs is to extract more units of independent work than the number of workers that can execute simultaneously such that there is enough work available for dynamic load balancing.
For prior work, this would mean starting at a lower level in the search tree where there are many more sub-trees available than the number of thread blocks that can execute simultaneously.
However, we have seen that starting at a deeper level in the tree could result in higher grid launch and memory overhead in one approach~\cite{kabbara2013parallel} or more redundant work in another approach~\cite{abu2018accelerating}.

\subsection{Challenge \#3: Memory as a Limiting Factor to Parallelism}\label{sec:challenge-memory}

GPUs have a relatively small memory capacity (compared to CPUs) while at the same time placing higher pressure on the memory capacity because of their massive parallelism.
Memory capacity on the GPU can be a limiting factor for parallelism in two ways.
First, the global memory capacity can limit the number of threads or thread blocks that can run simultaneously on the GPU if these threads or thread blocks require a large amount of global memory to store intermediate execution state.
Second, the shared memory capacity per streaming multiprocessor (SM) can limit the occupancy of threads on the SM if a large amount of shared memory is required to store frequently accessed data.
Both of these limits are encountered when traversing the vertex cover search tree.

For a thread block to traverse a sub-tree of the vertex cover search tree, it needs to manage a stack that stores the intermediate graph (a sub-graph with the solution vertices removed) at each level of the sub-tree.
An explicitly managed stack is used instead of recursion because different threads in a block all need to access the same intermediate graph.
Due to the expensive nature of dynamic memory allocation on GPUs, it is inefficient to grow the memory allocated for this stack dynamically.
Instead, a stack for each thread block is pre-allocated in global memory and provisioned for the maximum possible depth that the tree can reach.
Since GPUs execute many thread blocks at a time, enough memory needs to be available to support all the maximally-provisioned stacks for all the thread blocks simultaneously.
These stacks grow with the size of the graph, which can make the global memory capacity a limiting factor for parallelism for large graphs.

When a thread block visits a node in its sub-tree, it frequently accesses the intermediate graph at that node.
As an optimization, the intermediate graph can be placed in shared memory for fast access.
However, placing the intermediate graph in shared memory can make the shared memory capacity a limiting factor for occupancy for large graphs.
These global and shared memory capacity limits make efficient memory management an important challenge when parallelizing the traversal of the vertex cover search tree on GPUs.

\section{Parallelizing Vertex Cover on GPUs}

\subsection{Hybrid Traversal Approach using a Global Worklist}

One way to address the challenges mentioned in Section~\ref{sec:challenges} is by using a global worklist.
Rather than assigning thread blocks to entire sub-trees, a thread block can be assigned to a single node in the tree.
Upon branching, the thread block could add its node’s children to a global worklist where other thread blocks can pick them up and process them.
This approach substantially increases the amount of parallelism extracted from the traversal because it treats each tree node as a unit of parallelism as opposed to entire sub-trees.
It also substantially reduces load imbalance because tree nodes are more similar to each other in load than entire sub-trees, and there is more of them to go around for dynamic load balancing.
Additionally, a global worklist obviates the need for each thread block to maintain a local stack, which reduces the pressure on the global memory capacity.
However, using a global worklist has two major drawbacks.
The first drawback is that it converts the depth-first traversal of the search tree into a breadth-first traversal, which results in an exponential explosion of the number of tree nodes that need to be added to the global worklist, quickly exceeding the worklist's capacity.
The second drawback is that it creates a high amount of contention between thread blocks when accessing the global worklist, which becomes a serialization point in the program.

To reap the benefits of global worklists while mitigating their drawbacks, we propose a hybrid approach.
In our approach, each thread block traverses a sub-tree in depth-first order while keeping track of its intermediate state using a local stack (stored in global memory).
However, every time a thread block branches from a tree node, it first checks the global worklist.
If the number of entries in the global worklist is below a certain threshold, the thread block will add one of its node's children to the global worklist and move to process the other child.
Otherwise, if the number of entries in the global worklist exceeds the threshold, the thread block will push its node’s child to its local stack and move to process the other child.
When the thread block reaches the bottom of the tree and needs to find more work to do, it first attempts to pop a tree node from its local stack.
If the local stack is empty, the thread block will then take a tree node from the global worklist and begin to traverse the sub-tree rooted at that node.

This hybrid approach of using both a global worklist and per-block local stacks captures the advantages of the two individual approaches.
On the one hand, using a global worklist helps extract more parallelism from the computation and performs better dynamic load balancing.
On the other hand, using the local stacks when the global worklist is sufficiently full prevents the exponential explosion of the number of tree nodes to be added to the global worklist.
It also reduces contention on the global worklist since thread blocks will only add to the global worklist if it contains a small number of elements, and will always try to remove work from their local stacks before trying to remove from the global worklist.

\begin{figure}[t]
    \small
    \include{fig/2-background/parallel-mvc}
    \vspace{-10pt}
    \caption{Minimum Vertex Cover using a Hybrid Approach with a Global Worklist and Per-block Local Stacks}\label{fig:parallel-mvc}
\end{figure}

Figure~\ref{fig:parallel-mvc} shows the pseudocode for {\sc Minimum Vertex Cover} using our hybrid approach with a global worklist and per-block local stacks.
Initially, all stacks are empty and the global worklist contains the root node of the tree.
Each thread block first tries to pop work from its local stack (lines 5-6).
If unsuccessful, the thread block tries to get work from the global worklist (lines 7-8).
If the worklist indicates that the traversal is done (see Section~\ref{sec:global-worklist}), then the block terminates (lines 9-10).
If the block is successful at obtaining a tree node from its local stack or the global worklist, then it starts by reducing the intermediate graph at that tree node (line 11).
Next, it checks the stopping condition to see if it is still possible to find a solution on this branch (line 12, see Section~\ref{sec:background-alg}).
If not possible, it sets a flag to obtain a new tree node from the stack or the global worklist on the next iteration.
If it is still possible to find a solution, the block checks if it has already found a solution at the current tree node (line 17), and if yes, it atomically updates the current best solution size (line 18).
It also sets a flag to obtain a new tree node from the stack or the global worklist on the next iteration (line 19).
If the block has not found a solution yet then it needs to branch (line 20).
The block sets up one of the child nodes by removing the neighbors of the max-degree vertex from the graph and adding them to the vertex cover (lines 21-22).
If the worklist is sufficiently full (line 23), the child is pushed to the stack (line 24), otherwise it is added to the worklist (lines 25-26).
The thread block then sets up the other child node by removing just the max-degree vertex from the graph and adding it to the vertex cover (lines 27-28).
The block will proceed to process this child node on the next iteration, so it sets a flag that it does not need to obtain a new tree node from the stack or the global worklist on the next iteration.

For space constraints, we omit the pseudocode for PVC which is largely similar to that of MVC in Figure~\ref{fig:parallel-mvc} with a few differences.
The differences in the stopping condition are already described in Section~\ref{sec:background-alg}.
Another difference is that when we find a vertex cover whose size does not exceed $k$, rather than updating best and continuing, we set a flag telling other blocks that a vertex cover has been found and we terminate.
We also add a condition at the beginning of the loop (before line 4) where blocks check the flag and terminate if a vertex cover has been found before picking up another tree node to work on.

With this hybrid approach, we have tackled the first two challenges described in Section~\ref{sec:challenges}, namely extracting massive parallelism and load imbalance.
However, we are still using a local stack per thread block so the third challenge of memory as a limiting factor to parallelism remains.
We address this challenge in upcoming sections.

\subsection{Graph Representation and Operations}\label{sec:graph-representation}

The original graph is represented using the commonly used Compressed Sparse Row (CSR) format~\cite{bell2008efficient}.
This representation is compact, requiring $\mathcal{O}(|V|+|E|)$ memory.
It also makes it easy to access the incident edges and neighbors of a given vertex.
A single copy of the CSR graph representation exists and is accessed by all thread blocks and never modified.

The intermediate graphs which are stored in the local stacks and the global worklist are not represented using CSR.
One reason is that the CSR format is expensive to modify which makes it inconvenient for removing vertices and edges from the graph.
Another reason is that the CSR format would make the local stacks and global worklist require too much memory, even with $\mathcal{O}(|V|+|E|)$ memory consumption.
Recall from Section~\ref{sec:challenge-memory} that efficient memory management is critical to be able to support large graphs.

Instead of using CSR, the intermediate graphs along with the set of removed vertices ($(G, S)$ in Figure~\ref{fig:parallel-mvc}) are jointly represented using just a \textit{degree array}.
The degree array is an array with one element per original vertex that stores the degree of the vertex if the vertex is still in the graph, or a sentinel value if the vertex has been removed from the graph and added to the solution $|S|$.
The degree array representation has been used to represent intermediate graphs when searching for vertex covers~\cite{DBLP:conf/faw/Abu-KhzamLMN10,kabbara2013parallel}.
It is particularly useful in our implementation for two reasons.
The first reason is that the array only consumes $\mathcal{O}(|V|)$ memory which limits the memory consumption of the local stacks and global worklist.
The second reason is that when combined with the original graph, it is sufficient to represent the updated graph without any other information.
This property is important for dynamic load balancing because we need to be able to put $(G, S)$ in the global worklist where any other thread block can pick it up, so $(G, S)$ must be self-contained.

Operations on the intermediate graph in Figure~\ref{fig:parallel-mvc} are performed in parallel via collaboration between the threads within the block.
Applying the reduction rules (line 11) is discussed separately in Section~\ref{sec:reduction-rules}.
To find the vertex with maximum degree (line 16), a parallel reduction tree is performed on the degree array.
To remove a single vertex from the graph and add it to the solution (lines 27-28), the vertex's degree is set to a sentinel value by one thread and the vertex's neighbors are distributed across the threads to decrement their degrees in parallel.
To remove the neighbors of a single vertex from the graph and add them to the solution (lines 21-22), the vertex's neighbors are distributed across the threads.
For each neighbor, a thread will iterate over the neighbor's neighbors and atomically decrement their degrees, then set the degree of the neighbor to a sentinel value.
To find the number of vertices in the solution ($|S|$ on lines 12 and 18), a reduction tree could be performed over the degree array to count the number of sentinel values.
However, as an optimization, we store an additional counter with the degree array that tracks the number of deleted vertices, and update that counter whenever we delete a vertex.

\subsection{Implementation of the Global Worklist}\label{sec:global-worklist}

We implement the global worklist using the Broker Work Distributor (BWD)~\cite{kerbl2018broker} which is a state-of-the-art worklist data structure for dynamic work distribution.
We make one modification to the BWD data structure to support our algorithm.
By design, if the worklist is empty, BWD returns that it cannot remove any elements from the worklist.
The worklist is considered empty if all blocks that have added or committed to add an entry have corresponding blocks that have committed to remove an entry.
However, the worklist may be empty in one of two situations.
The first situation is where some blocks are still executing and may commit to add entries to the worklist in the future.
In this situation, we would like to keep checking the worklist until the new work arrives.
The second situation is where no blocks are executing and they are all trying to remove work from the empty worklist.
In this situation, we can expect that no blocks will commit to add work in the future, which means that the work is done and we can safely terminate.

To handle these two situations, we wrap the BWD function for removing worklist entries in a loop.
Each iteration of the loop first attempts to remove an entry from the worklist.
If it succeeds, we return this entry so the block can process it.
If it fails, we atomically check if the worklist is empty and if the number of thread blocks trying to remove from the worklist is all the thread blocks in the grid.
If the check succeeds, we return that we are done so the block can exit (lines 9-10 in Figure~\ref{fig:parallel-mvc}).
In the parameterized version, we also check the flag that indicates that a vertex cover has been found.
If the flag is set, then we return that we are done.
If we are not done, then we let the thread block sleep for some time then go back to the beginning of the loop and repeat the process.

\subsection{Reduction Rules}\label{sec:reduction-rules}

When a thread block visits a tree node, it applies the three reduction rules (degree-one, degree-two-triangle, and high-degree) described in Section~\ref{sec:background-alg} until the graph no longer changes.
Each rule is executed in parallel by all the threads in the block, so care must be exercised when implementing them in parallel.
For the degree-one rule, different threads find different degree-one vertices simultaneously.
However, different degree-one vertices may have a common neighbor so care is exercised to ensure that the neighbor is removed only once.
Moreover, two threads may simultaneously find two degree-one vertices that are neighbors of each others, so only one of the two vertices is removed (the one with the smaller vertex ID), not both.
For the degree-two-triangle rule, different threads find different degree-two vertices simultaneously and check if they are part of a triangle.
However, different degree-two vertices may participate in the same triangle, so the neighbors of only one of these vertices (the one with the smaller vertex ID) are removed.
We handle all these cases in our parallel implementation of the reduction rules.

\begin{table*}[t]
    \centering
    \caption{Execution Time (in Seconds)}\label{tab:time}
    \vspace{-5pt}
    \resizebox{\textwidth}{!}{
        
\begin{tabular}{l|l|r|r|r|r|r|r|r|r|r|r|r|r|r|r|r|}
    \cline{2-17}
    \multicolumn{1}{l|}{\multirow{3}{*}{\textbf{}}} & \multicolumn{1}{c|}{\multirow{3}{*}{\textbf{Graph}}} & \multicolumn{1}{c|}{\multirow{3}{*}{\textbf{$|V|$}}} & \multicolumn{1}{c|}{\multirow{3}{*}{\textbf{$|E|$}}} & \multicolumn{1}{c|}{\multirow{3}{*}{\textbf{$\frac{|E|}{|V|}$}}} & \multicolumn{3}{c|}{\multirow{2}{*}{\textbf{MVC}}} & \multicolumn{9}{c|}{\textbf{PVC}} \\
    \cline{9-17}
     & & & & & \multicolumn{3}{c|}{ } & \multicolumn{3}{c|}{\textbf{\textit{k = min -- 1}}} & \multicolumn{3}{c|}{\textbf{\textit{k = min}}} & \multicolumn{3}{c|}{\textbf{\textit{k = min + 1}}} \\
    \cline{6-17}
     & & & & & \textbf{Sequential} & \textbf{StackOnly} & \multicolumn{1}{c|}{\textbf{Hybrid}} & \textbf{Sequential} & \textbf{StackOnly} & \multicolumn{1}{c|}{\textbf{Hybrid}} & \textbf{Sequential} & \textbf{StackOnly} & \textbf{Hybrid} & \textbf{Sequential} & \textbf{StackOnly} & \textbf{Hybrid}   \\
    \cline{2-17}
    \multicolumn{1}{l|}{\multirow{10}{*}{\rotatebox[origin=c]{90}{High degree}}}
    & p\_hat\_300\_1~\cite{johnson1996cliques} & 300 & 33917 & 113 & 0.138 & 0.780 & \textbf{0.021} & 0.140 & 0.782 & \textbf{0.023} & 0.031 & 0.021 & \textbf{0.016} & 0.028 & \textbf{0.001} & 0.003 \\
    & p\_hat\_300\_2~\cite{johnson1996cliques} & 300 & 22922 & 76 & 1.262 & 15.681 & \textbf{0.029} & 1.266 & 15.714 & \textbf{0.029} & 0.016 & 0.021 & \textbf{0.016} & 0.016 & \textbf{0.010} & 0.016 \\
    & p\_hat\_300\_3~\cite{johnson1996cliques} & 300 & 11460 & 38 & 200.990 & 2,197.583 & \textbf{1.657} & 193.597 & 2,199.056 & \textbf{1.658} & \textbf{0.047} & 0.528 & 0.056 & 0.006 & \textbf{0.005} & 0.008 \\
    & p\_hat\_500\_1~\cite{johnson1996cliques} & 500 & 93181 & 186 & 1.150 & 7.787 & \textbf{0.092} & 1.456 & 7.823 & \textbf{0.090} & 0.146 & 0.145 & \textbf{0.019} & 0.139 & \textbf{0.006} & 0.007 \\
    & p\_hat\_500\_2~\cite{johnson1996cliques} & 500 & 61804 & 124 & 102.553 & 1,541.506 & \textbf{1.558} & 100.602 & 1,542.344 & \textbf{1.559} & \textbf{0.069} & 0.122 & 0.101 & 0.072 & \textbf{0.036} & 0.070 \\
    & p\_hat\_500\_3~\cite{johnson1996cliques} & 500 & 30950 & 62 & $>$ 2 hrs & $>$ 2 hrs & \textbf{1,018.898} & $>$ 2 hrs & $>$ 2 hrs & \textbf{1,027.504} & \textbf{2.480} & 928.941 & 25.636 & \textbf{0.022} & 0.083 & 0.095 \\
    & p\_hat\_700\_1~\cite{johnson1996cliques} & 700 & 183651 & 262 & 4.838 & 31.245 & \textbf{0.238} & 8.054 & 31.200 & \textbf{0.178} & 0.672 & 0.584 & \textbf{0.188} & 0.409 & 0.593 & \textbf{0.075} \\
    & p\_hat\_700\_2~\cite{johnson1996cliques} & 700 & 122922 & 176 & 1,949.591 & $>$ 2 hrs & \textbf{31.241} & 1,833.827 & $>$ 2 hrs & \textbf{31.507} & 2.903 & 42.947 & \textbf{0.243} & 0.221 & \textbf{0.060} & 0.074 \\
    & p\_hat\_1000\_1~\cite{johnson1996cliques} & 1000 & 377247 & 377 & 58.056 & 495.296 & \textbf{1.400} & 63.104 & 495.099 & \textbf{1.397} & 1.456 & 5.099 & \textbf{0.135} & 1.151 & 0.043 & \textbf{0.017} \\
    & p\_hat\_1000\_2~\cite{johnson1996cliques} & 1000 & 254701 & 255 & $>$ 2 hrs & $>$ 2 hrs & \textbf{4,527.601} & $>$ 2 hrs & $>$ 2 hrs & \textbf{4,596.877} & \textbf{1.263} & 8.128 & 4.099 & \textbf{0.627} & 0.902 & 0.939 \\
    & movielens-100k\_rating~\cite{kunegis2013konect} & 2625 & 94834 & 36 & 4.906 & \textbf{0.115} & 0.132 & 4.840 & \textbf{0.114} & 0.133 & \textbf{0.012} & 0.019 & 0.023 & \textbf{0.012} & 0.019 & 0.023 \\
    & wikipedia\_link\_lo~\cite{kunegis2013konect} & 3811 & 83029 & 22 & $>$ 2 hrs & $>$ 2 hrs & \textbf{387.628} & $>$ 2 hrs & $>$ 2 hrs & \textbf{421.803} & $>$ 2 hrs & \textbf{0.031} & 0.045 & \textbf{0.020} & 0.030 & 0.047 \\
    & wikipedia\_link\_csb~\cite{kunegis2013konect} & 5561 & 187269 & 34 & 0.372 & 39.227 & \textbf{0.034} & 0.151 & 39.147 & \textbf{0.035} & 0.158 & 0.007 & \textbf{0.007} & 0.107 & 0.007 & \textbf{0.006} \\
    \cline{2-17}
    \multicolumn{1}{l|}{\multirow{5}{*}{\rotatebox[origin=c]{90}{Low degree}}}
    & US power grid~\cite{kunegis2013konect} & 4942 & 6594 & 1.33 & 145.574 & 1.518 & \textbf{0.852} & 141.734 & 1.531 & \textbf{0.853} & 0.002 & 0.001 & \textbf{0.001} & 0.002 & 0.001 & \textbf{0.001} \\
    & LastFM Asia~\cite{snap} & 7624 & 27806 & 3.65 & 83.389 & 4.345 & \textbf{0.939} & 81.894 & 4.395 & \textbf{1.052} & 0.005 & 0.009 & \textbf{0.005} & 0.005 & 0.008 & \textbf{0.005} \\
    & Sister Cities~\cite{kunegis2013konect} & 14275 & 20573 & 1.44 & 5.634 & 2.850 & \textbf{0.106} & 5.526 & 2.853 & \textbf{0.116} & 0.004 & 0.005 & \textbf{0.002} & 0.004 & 0.005 & \textbf{0.003} \\
    & vc-exact\_023~\cite{dzulfikar2019pace} & 27718 & 133665 & 4.82 & $>$ 2 hrs & $>$ 2 hrs & $>$2 hrs & $>$ 2 hrs & $>$ 2 hrs & $>$ 2 hrs & 1.539 & 0.898 & \textbf{0.878} & 1.537 & 0.898 & \textbf{0.881} \\
    & vc-exact\_009~\cite{dzulfikar2019pace} & 38453 & 174645 & 4.54 & $>$ 2 hrs & $>$ 2 hrs & $>$ 2 hrs & $>$ 2 hrs & $>$ 2 hrs & $>$ 2 hrs & 2.883 & \textbf{1.605} & 1.651 & 2.878 & \textbf{1.605} & 1.642 \\
    \cline{2-17}
\end{tabular}

    }
\end{table*}

\subsection{Memory Management}\label{sec:memory-management}

Recall from Section~\ref{sec:challenge-memory} that efficient memory management is critical for supporting large graphs.
The total amount of global memory needed for storing all the per-block local stacks is dependent on three factors: the size of a stack entry (i.e., size of the intermediate graph), the number of stack entries per stack (i.e., maximum depth of the search tree), and the number of stacks (i.e., number of thread blocks).
We have already seen in Section~\ref{sec:graph-representation} that we limit the size of a stack entry by representing the intermediate graph using a degree array which requires $\mathcal{O}(|V|)$ space.
To limit the number of stack entries per stack, we run the greedy algorithm to approximate the minimum vertex cover on the CPU (see Section~\ref{sec:background-alg}) and use the approximation as the limit on the stack depth when starting the GPU kernel since no thread block will ever go deeper in the tree than the size of this minimum.
In the parameterized version, the parameter \textit{k} is used as the bound.
To limit the number of stacks, we must limit the number of thread blocks as the size of the graph gets larger.
To limit the number of thread blocks while maintaining the total number of threads needed to achieve the highest device occupancy, we must use a large number of threads per block.
Hence, the number of threads per block must be carefully selected based on the size of the graph to ensure that the highest device occupancy is achieved for large graphs.

As for shared memory, shared memory is primarily used for each thread block to store the intermediate graph for the tree node it is currently working on.
The total amount of shared memory needed per SM depends on two factors: the amount of shared memory needed per block (i.e., the size of the intermediate graph) and the number of blocks per SM.
We have already seen that we limit the size of the intermediate graph by using a degree array to represent it which requires $\mathcal{O}(|V|)$ space.
To limit the number of thread blocks per SM while maintaining the total threads needed to achieve maximum SM occupancy, we must use a large number of threads per block.
Hence, the choice of the number of threads per block not only considers the impact of the global memory capacity on the number of blocks that can run concurrently on the device, but also the impact of the shared memory capacity on the number of blocks that can run concurrently per SM.

To satisfy these constraints while maximizing occupancy, we select then number of threads per block as follows.
We determine an upper-limit on the number of threads per block based on the hardware limit on the block size and $|V(G)|$, whichever is smaller.
We use $|V(G)|$ as an upper-limit because it is not useful to have more threads in the block than the number of vertices in the graph because these threads will not perform any work.
We also determine a lower-limit on the number of threads per block based on the desired number of threads to achieve full occupancy and the upper-limit on number of blocks that can run simultaneously.
The upper-limit on the number of blocks is the minimum of the following limits: the hardware limit on the number of simultaneous blocks, the shared memory limit on the number of simultaneous blocks, and the global memory limit on number of simultaneous blocks (i.e., number of stacks that can be stored).
If the lower-limit is less than the upper-limit, we select a thread block size within the range that is a power of two.
If the lower-limit is greater than the upper-limit, then it is impossible to achieve full occupancy.
In this case, we select the upper-limit as the thread block size and let the kernel execute without achieving full occupancy.
In practice, the shared memory capacity tends to be more restrictive than the global memory capacity for most graphs.
For this reason, we provide two versions of each kernel, one that uses shared memory to store the intermediate graph that the block is currently working on, and one that uses global memory to store the intermediate graph.
If the lower-limit is too high because of the shared memory constraint, we relax the shared memory constraint by falling back on the kernel that uses global memory to store the intermediate graph.

\section{Evaluation}

\subsection{Methodology}\label{sec:methodology}

We implement and evaluate three different code versions:
\begin{itemize}
    \item \textit{Sequential}: This implementation executes on a single CPU thread. The objective of evaluating this implementation is just for reference. A fair comparison to CPUs would entail comparing to a parallel CPU implementation, but this is not the aim of our work. Our work aims to show how the vertex cover search tree traversal can be parallelized using GPUs.
    \item \textit{StackOnly}: This implementation parallelizes sub-trees starting at a specific level across thread blocks. Each thread block makes its way down to its sub-tree from the root then proceeds to traverse its sub-tree using a per-block local stack, similar to what prior work does~\cite{abu2018accelerating}.
    \item \textit{Hybrid}: This implementation uses the hybrid approach that leverages per-block local stacks as well as a global worklist to assist with dynamic load balancing.
\end{itemize}
For a fair comparison, all the versions use the same data structure, apply the same reduction rules, and use the same strategy to compute an approximate minimum on the CPU before traversing the search tree.

We implement our code using C++ and CUDA.
The CPU implementation is evaluated on an AMD EPYC 7551P CPU with 64GB of main memory.
The GPU implementations are evaluated on a Volta V100 GPU with 32GB of device memory.

For the \textit{StackOnly} and \textit{Hybrid} implementations, we follow the strategy described in Section~\ref{sec:memory-management} to select between the shared memory and global memory kernels and to select the number of threads per block (block size).
The shared memory kernel is selected for all high-degree graphs (small $|V|$) and the global memory kernel is selected for all low-degree graphs (large $|V|$).
If multiple block sizes are possible, we try them all and report the best result.
However, one can still obtain performance benefits without selecting the best block size.
Sub-optimal selection of the block size would cause a geometric mean slowdown of 1.55$\times$ in the average case and 2.40$\times$ in the worst case for the StackOnly implementation, and 1.39$\times$ in the average case and 1.80$\times$ in the worst case for the Hybrid implementation.
Hence, the Hybrid implementation is more robust than the StackOnly implementation to a sub-optimal selection of block size, and the slowdown of a sub-optimal selection is within the speedup margins reported in our evaluation.

For the StackOnly implementation, to select the starting depth, we try three different depth values (8, 12, and 16) and report the best result.
Sub-optimal selection of the starting depth would result in a geometric mean slowdown of 1.18$\times$ in the average case and 1.37$\times$ in the worst case.

For the Hybrid implementation, we try global worklist sizes of $128K$, $256K$, and $512K$ entries, and threshold values of 0.25$\times$, 0.5$\times$, 0.75$\times$, and 1.0$\times$ the worklist size and report the best result.
Sub-optimal selection of the worklist size and threshold would result in a geometric mean slowdown of 1.18$\times$ in the average case and 1.32$\times$ in the worst case, which is within the speedup margins reported in our evaluation.

\subsection{Performance}\label{sec:eval-performance}

Table~\ref{tab:time} shows the execution time of each implementation for four different instances of the problem (MVC, PVC with $k=min-1$, PVC with $k=min$, PVC with $k=min+1$) across a wide range of graphs obtained from popular collections~\cite{johnson1996cliques,kunegis2013konect,dzulfikar2019pace,snap}.
We take the edge complements of graphs in the DIMACS collection~\cite{johnson1996cliques} like in prior work~\cite{abu2018accelerating}.
Table~\ref{tab:speedup} shows the speedup of the Hybrid implementation over the StackOnly implementation and the CPU implementation (Sequential) for the four instances, aggregated across two categories of graphs: graphs with high average degree (denoted as high-degree) and graphs with low average degree (denoted as low-degree).
Based on these results, we make three key observations.

\begin{table}[t]
    \centering
    \caption{Aggregate Speedup (Geometric Mean)}\label{tab:speedup}
    \vspace{-5pt}
    \resizebox{1.0\columnwidth}{!}{
        
\begin{tabular}{|l|r|r|r|r|r|r|r|r|}
    \hline
    \multicolumn{1}{|c|}{\multirow{3}{*}{\textbf{Category}}} & \multicolumn{4}{c|}{\textbf{Speedup of Hybrid over StackOnly}} & \multicolumn{4}{c|}{\textbf{Speedup of Hybrid over Sequential}} \\
    \cline{2-9}
     & \multirow{2}{*}{\textbf{MVC}} & \multicolumn{3}{c|}{\textbf{PVC}} & \multirow{2}{*}{\textbf{MVC}} & \multicolumn{3}{c|}{\textbf{PVC}}   \\
    \cline{3-5}\cline{7-9}
     &  & \textbf{\textit{k=min--1}} & \textbf{\textit{k=min}} & \textbf{\textit{k=min+1}} &  & \textbf{\textit{k=min--1}} & \textbf{\textit{k=min}} & \textbf{\textit{k=min+1}} \\
    \hline
    \textbf{High-degree} & 167.1$\times$ & 171.3$\times$ & 4.2$\times$ & 0.9$\times$ & 30.0$\times$ & 30.1$\times$ & 1.8$\times$ & 2.4$\times$ \\
    \hline
    \textbf{Low-degree} & 6.1$\times$ & 5.7$\times$ & 1.2$\times$ & 1.2$\times$ & 93.1$\times$ & 85.0$\times$ & 1.5$\times$ & 1.5$\times$ \\
    \hline
    \textbf{Overall} & 72.9$\times$ & 73.1$\times$ & 3.0$\times$ & 1.0$\times$ & 39.0$\times$ & 38.2$\times$ & 1.7$\times$ & 2.1$\times$ \\
    \hline
\end{tabular}

    }
\end{table}

The first observation is that the Hybrid implementation substantially outperforms the StackOnly implementation on high-degree graphs, while having moderate performance advantage on low-degree graphs.
Recall from Section~\ref{sec:challenge-balance} that the vertex cover search tree is imbalanced because whenever the tree branches, one branch removes a single high-degree vertex from the graph whereas the other branch removes all the neighbors of that high-degree vertex.
The higher the average degree of the graph, the higher the disparity in how many vertices are removed by each branch on average, and therefore, the more imbalanced the search tree is likely to be.
Since high-degree graphs are likely to have more imbalanced search trees, they are likely to  benefit more from the load balancing that the Hybrid implementation provides.

The second observation is that the Hybrid implementation substantially outperforms the StackOnly implementation on the difficult instances with long run-times (MVC and PVC with $k=min-1$), while having comparable performance on the easier instances with short run-times (PVC with $k=min$ and $k=min + 1$).
PVC with $k=min$ and PVC with $k=min+1$ stop as soon as a solution is found on one branch of their search trees, whereas MVC searches all branches of its tree to find the smallest vertex cover and PVC with $k=min-1$ searches all branches of its search tree without finding any solution.
Because MVC and PVC with $k=min-1$ search their trees more exhaustively, they are more likely to run into deeper branches that cause load imbalance, which makes them more likely to benefit from the load balancing that the Hybrid implementation provides.

We look further into the first and second observation in Section~\ref{sec:eval-load-balance} where we analyze the load balance of each implementation more thoroughly on different graphs and for different instances.
As shown in Table~\ref{tab:speedup}, the Hybrid implementation is faster than the StackOnly implementation on high-degree graphs by 167.1$\times$ for MVC and 171.3$\times$ for PVC with $k=min-1$.
Although the StackOnly implementation does outperform the Hybrid implementation in a few instances for select graphs as shown in Table~\ref{tab:time}, these cases tend to be on easier instances with short run-times and the performance difference is not usually significant.
For this reason, we are not motivated to design a criteria for selecting between the two GPU implementations.

Our third observation is that the Hybrid GPU implementation outperforms the Sequential CPU implementation substantially, especially for difficult instances with long run-times (MVC and PVC with $k=min-1$).
As mentioned in Section~\ref{sec:methodology}, a fair comparison to CPUs would entail comparing to a parallel CPU implementation.
We compare to Sequential just for reference to show that GPUs can have competitive performance compared to CPUs in this tree traversal algorithm that is normally considered difficult to parallelize.

\subsection{Load Balance}\label{sec:eval-load-balance}

\begin{figure}
    \centering
    \includegraphics[width=\columnwidth]{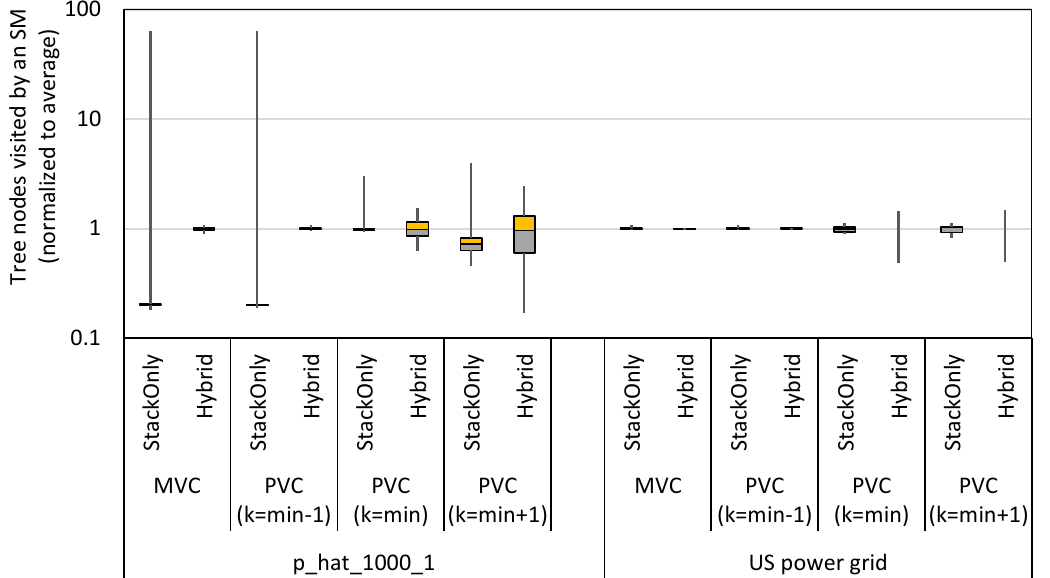}
    \caption{Distribution of Load across SMs}\label{fig:imbalance}
\end{figure}

Figure~\ref{fig:imbalance} compares the load distribution achieved by each of the StackOnly and Hybrid implementations for the four instances of the problem on two sample graphs.
Load is measured as the ratio of the number of tree nodes visited by an SM to the average number of tree nodes visited across all SMs.
The graphs picked are those at the two extremes, having the highest average degree (p\_hat\_1000\_1) and the lowest average degree (US power grid).
We make three key observations.

The first observation is that the StackOnly implementation has substantially higher load imbalance on the high-degree graph than on the low-degree graph.
The second observation is that the StackOnly implementation has substantially higher load imbalance on the difficult instances with long run-times (MVC and PVC with $k=min-1$) than on the easier instances with short run-times (PVC with $k=min$ and $k=min+1$).
These observations are consistent with the points mentioned in Section~\ref{sec:eval-performance} that high-degree graphs and difficult long-running instances are likely to suffer from more load imbalance.

The third observation is that the Hybrid implementation achieves better load balance than the StackOnly implementation.
For example, when the StackOnly implementation solves MVC on p\_hat\_1000\_1, more than 75\% of the SMs take less than 0.21$\times$ the average load, whereas one SM takes 63.98$\times$ the average load.
In contrast, when the Hybrid implementation solves the same instance, the least loaded SM takes 0.89$\times$ the average load whereas the most loaded SM takes 1.07$\times$ the average load.
These results demonstrate the effectiveness of the Hybrid implementation at achieving load balance.

\subsection{Breakdown of Execution Time}\label{sec:breakdown}

\begin{figure}
    \centering
    \includegraphics[width=\columnwidth]{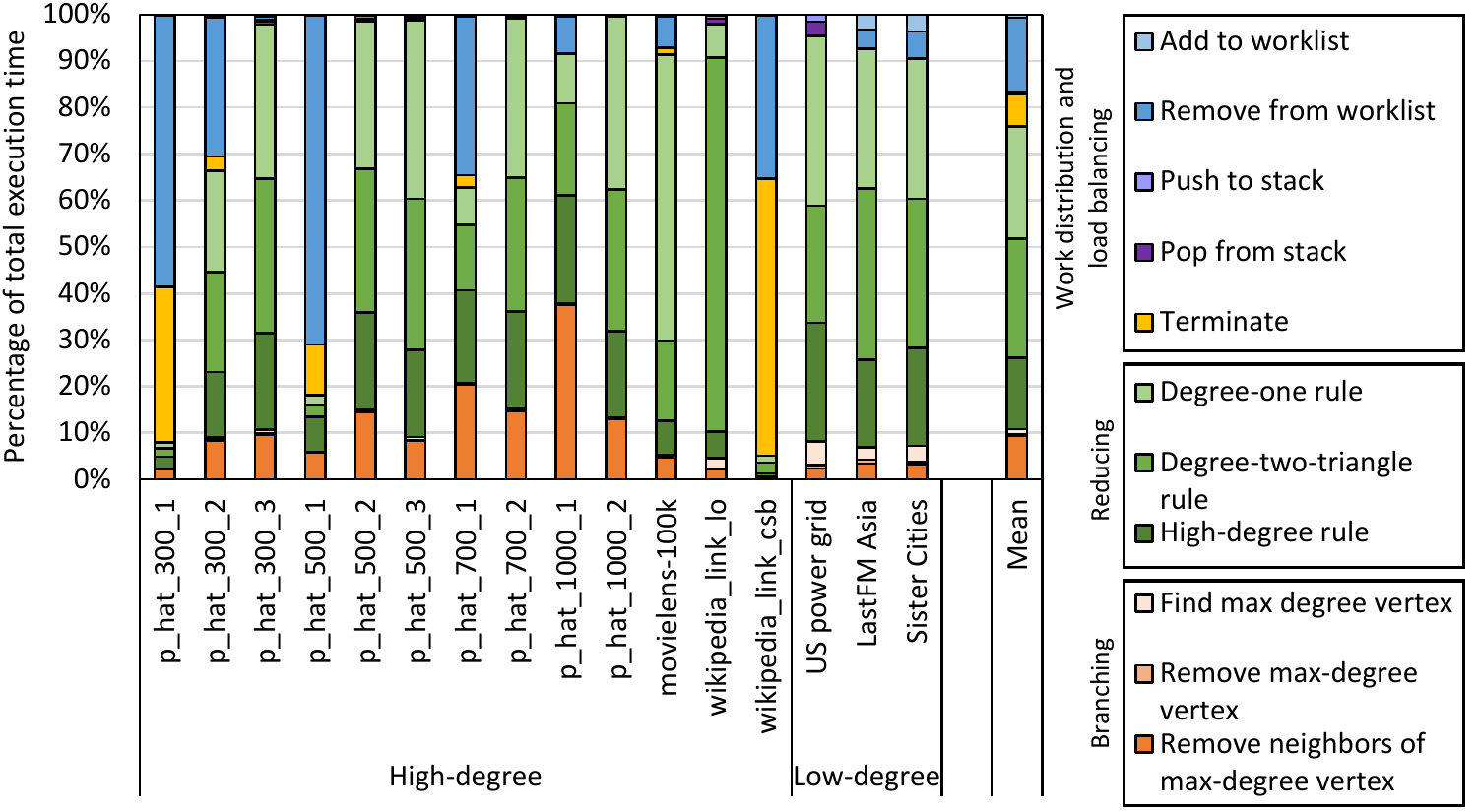}
    \caption{Breakdown of Execution Time for MVC}\label{fig:counters}
\end{figure}

To better understand where our traversal code is spending time, Figure~\ref{fig:counters} shows the breakdown of the execution time of our MVC kernel for different graphs.
To measure this breakdown, we instrument our code to use the SM clocks to get the number of cycles spent by each thread block on each activity.
We then normalize the cycle counts to the total number of cycles executed by the thread block and take the mean across all thread blocks.
We make three key observations.

The first observation is that the kernel spends 24.1\% of its time on average on activities related to work distribution and load balancing, of which 16.0\% is spent removing from the global worklist.
Removing from the worklist is expensive because there may be high contention on the worklist from other blocks, or because the worklist may be empty requiring the block to wait for new work to arrive (see Section~\ref{sec:global-worklist}).
This overhead is acceptable given the difficulty of extracting parallelism and load balancing in this particular problem.

The second observation is that the kernel spends 65.2\% of its time (the majority of its time) on average on the reduction rules.
This time is well-spent because the reduction rules allow the kernel to make the fastest progress towards finding a solution.
The time is distributed almost evenly across the different rules.

The third observation is that the kernel spends 10.7\% of its time on average on activities related to branching, of which 9.4\% is spent on removing the neighbors of the max-degree vertex.
It is noteworthy that removing the neighbors of the max-degree vertex takes less time in low-degree graphs than in high-degree graphs since there are fewer neighbors to remove on average in the low-degree graphs.

\subsection{Comparison with Prior Work}

Table~\ref{tab:time-related} compares the performance of our implementation to that of the most recent prior work for GPUs~\cite{abu2018accelerating}.
This prior work uses an approach similar to our StackOnly approach which distributes sub-trees starting at a specific level across thread blocks then has each block make its way down to its sub-tree from the root and traverse its sub-tree using a per-block local stack.
In particular, it solves the PVC instance and evaluates using $k=min$.
The times reported in the paper~\cite{abu2018accelerating} are used directly for comparison and are replicated in Table~\ref{tab:time-related}.
We note that this comparison is not fair because prior work uses two AMD FirePro D500 GPUs with 3GB of memory each, while we use a more powerful Volta V100 GPU with 32GB of memory.
However, we are unable to evaluate their performance on our system because the code is not publicly available.
The objective of this comparison is to show that our approach is highly competitive with prior GPU solutions for the vertex cover problem.

\begin{table}[t]
    \centering
    \caption{Comparison of Execution Time (in Seconds) with Prior Work}\label{tab:time-related}
    \resizebox{0.9\columnwidth}{!}{
        
\begin{tabular}{|l|r|r|r|r|}
    \hline
    \textbf{Graph} & \textbf{Sequential} & \textbf{StackOnly} & \textbf{Hybrid} & \textbf{Abu Khuzam et al.~\cite{abu2018accelerating}} \\
    \hline
    p\_hat\_300\_1 & 0.031 & 0.021 & \textbf{0.016} & 4.400      \\
    p\_hat\_300\_2 & 0.016 & 0.021 & \textbf{0.016} & 5.000      \\
    p\_hat\_300\_3 & \textbf{0.047} & 0.528 & 0.056 & 2.800      \\
    p\_hat\_500\_1 & 0.146 & 0.145 & \textbf{0.019} & 10.700      \\
    p\_hat\_500\_2 & \textbf{0.069} & 0.122 & 0.101 & 10.100      \\
    p\_hat\_500\_3 & \textbf{2.480} & 928.941 & 25.636 & 6.000      \\
    p\_hat\_700\_1 & 0.672 & 0.584 & \textbf{0.188} & 21.000      \\
    p\_hat\_700\_2 & 2.903 & 42.947 & \textbf{0.243} & 14.800      \\
    p\_hat\_1000\_1 & 1.456 & 5.099 & \textbf{0.135} & 48.300      \\
    p\_hat\_1000\_2 & \textbf{1.263} & 8.128 & 4.099 & 30.800      \\
    \hline
\end{tabular}
    }
\end{table}

\section{Related Work}

In the last few decades, a lot of effort has been devoted to developing fast and simple exact algorithms for NP-hard problems~\cite{DBLP:books/fm/GareyJ79} and MVC is no exception.
One of the first examples is the $\mathcal{O}(2^{n/3})$-time algorithm of Tarjan and Trojanowski~\cite{DBLP:journals/siamcomp/TarjanT77} for {\sc Maximum Independent Set (MIS)} on $n$-vertex graphs.
Note that MIS is equivalent to MVC since the complement of a minimum vertex cover is a \emph{maximum independent set}, i.e., a maximum set of pairwise non-adjacent vertices.
The aforementioned $\mathcal{O}(2^{n/3})$-time  algorithm is significantly faster than the trivial $\mathcal{O}(2^{n})$-time brute-force algorithm. 
Considerable improvements were made in the algorithm of Robson~\cite{DBLP:journals/jal/Robson86, DBLP:journals/jal/RobsonNew} which runs in $\mathcal{O}(1.1889^n)$-time (further improvements are also known~\cite{DBLP:journals/iandc/XiaoN17,DBLP:journals/jco/XiaoN17}).

For PVC, when the size of the vertex cover we are looking for, denoted by $k$, is sufficiently smaller than $n$, much faster algorithms exist.
A problem is said to be \emph{fixed-parameter tractable (FPT)}, if it can be solved in time $f(k) \cdot n^{\mathcal{O}(1)}$, where $f$ only depends on $k$ (usually exponential in $k$) and $n$ is the size of input. 
In 1988, Fellows provided an $\mathcal{O}(2^k \cdot n)$ algorithm for PVC, showing that the problem is fixed-parameter tractable (a recent exposition can be found in Downey and Fellows~\cite{DF97}).
The algorithm is based on the bounded search tree technique discussed in Section~\ref{sec:background-alg}.
Since then, and after a long series of  works~\cite{DBLP:journals/ipl/BalasubramanianFR98,DBLP:conf/wg/ChenKJ99,DBLP:journals/networks/ChenLJ00,DBLP:conf/aaecc/FellowsK93,BussG93}, the asymptotic upper bound on the running time of PVC was improved to $\mathcal{O}(1.2738^k + kn)$ by Chen et al.~\cite{DBLP:journals/tcs/ChenKX10}.

There are many serial~\cite{DBLP:conf/siamcsc/HespeL0S20,DBLP:journals/tcs/AkibaI16,DBLP:conf/iwpec/DzulfikarFH19} and parallel~\cite{DBLP:journals/algorithmica/Abu-KhzamLSS06,DBLP:journals/jpdc/Abu-KhzamDMN15} implementations that solve the vertex cover problem on CPUs.
Our work focuses on solving the problem on GPUs, which has only recently gained attention.
Some recent works provide approximate/heuristic algorithms for MVC~\cite{toume2014gpu} and MIS~\cite{burtscher2018high,imanaga2020efficient} on GPUs.
The focus of our work is on the exact algorithms which follow the hard-to-parallelize branch-and-reduce paradigm.
Section~\ref{sec:challenges} already compares to prior works~\cite{abu2018accelerating,kabbara2013parallel} that parallelize exact vertex cover algorithms on GPUs.
These works distribute sub-trees starting at at the same level across thread blocks.
We show that our approach can achieve substantially better performance via improved load balancing.
Liu et al.~\cite{jufinding} traverse the top of the tree on the CPU and send sub-trees to the GPU whenever the size of the graph drops below a certain threshold.
This approach requires frequent communication between the CPU and the GPU, and results in launching many small grids (one single-block grid per sub-tree) which is known to underutilize device resources.

Search tree traversal on GPUs has been explored in the context of other problems.
For example, recent work has been done on graph pattern mining~\cite{chen2020pangolin,chen2021sandslash}, maximal clique enumeration~\cite{jenkins2011lessons,wei2021accelerating}, and $k$-clique counting~\cite{almasri2021k}.
These problems usually have a sufficient number of sub-trees available at the first or second level of the search tree such that distributing sub-trees across thread blocks can achieve adequate load balance.
The vertex cover problem is different in that the search tree is narrower and highly imbalanced, which makes extracting enough parallelism more difficult.
Other problems involving search tree traversal that have been solved on GPUs include the N-Queens problem~\cite{zhang2011optimization} and minimax tree search~\cite{rocki2009parallel}.
To extract enough parallelism, these approaches distribute sub-trees across threads or blocks starting at a certain depth in the tree, similar to what prior work~\cite{abu2018accelerating,kabbara2013parallel} does for the vertex cover problem.
Our work uses a global worklist to allow thread blocks to contribute branches of their sub-trees at any level for other idle blocks to process.
Another work on the N-Queens problem~\cite{plauth2015using} uses dynamic parallelism to parallelize the search tree traversal.
We avoid using dynamic parallelism in our implementation because it is known to be inefficient when many small grids are launched~\cite{el2016klap,olabi2022compiler}.

\section{Conclusion}

We present techniques for parallelizing exact algorithms for {\sc Minimum Vertex Cover} and {\sc Parameterized Vertex Cover} on GPUs.
We propose a hybrid approach for work distribution and dynamic load balancing where each thread block uses a local stack to traverse a sub-tree, but contributes branches of its sub-tree to a global worklist on an as-needed basis, extracting just enough parallelism for load balancing without incurring too much overhead.
We represent intermediate graphs using degree arrays to ensure that they are compact so that memory consumption does not limit parallelism, but at the same time self-contained so that they can be shared across different thread blocks in the load balancing process.
We implement CUDA kernels for solving both MVC and PVC using our proposed approach, and show that they achieve substantial performance and load balance improvements, especially on difficult instances of the problem and on graphs with high average degree.
Our implementations have been open sourced to enable further research on parallel solutions to the vertex cover problem and other similar problems involving parallel traversal of narrow and highly imbalanced search trees.

\balance
\bibliographystyle{IEEEtran}
\bibliography{main}

\end{document}